\documentclass[fleqn,twoside]{article} 
\usepackage{espcrc2}
\usepackage{graphicx} 
\title{Spacecraft Tracking with the SKA} 
\author{D.L.~Jones\address{Jet Propulsion Laboratory, California 
Institute of Technology, Pasadena, CA, USA, dj@sgra.jpl.nasa.gov} 
\thanks{This work was carried out at the Jet Propulsion Laboratory, 
California Institute of Technology, under contract with the 
National Aeronautics and Space Administration.}}

\begin{document}

\begin{abstract}
The possibility of using the SKA for occasional support of 
scientific space missions should be considered, along with 
all other science goals, in the design of the array.  The benefits of 
higher data rates from distant spacecraft during high priority, 
short duration mission phases can be dramatic, while the technical 
requirements on the SKA to allow this capability are very few. 
The most fundamental requirement is coverage of the primary  
downlink frequencies for deep space missions.   
Additional benefits can result from real-time high precision 
angular position measurements of spacecraft.  Such measurements 
allow spacecraft navigation with reduced errors and reduced risk, 
but require at least some long (thousands of km) 
baselines.   
\vspace{1pc}
\end{abstract}
\maketitle

\section{INTRODUCTION}

Why use the SKA to track spacecraft?  This is not generally 
considered to be a part of radio astronomy, after all.  However, 
there are multiple precedents for the temporary use of radio 
astronomy facilities to support scientific space missions, including 
the use of the VLA during Voyager's flyby of Neptune \cite{1} \cite{7}, 
and the combining of signals from the Parkes 64-m antenna with those from the 
Deep Space Network (DSN) station at Tidbinbilla \cite{2}.  In each of these cases, 
the goal was to increase the science return from a unique opportunity 
and a large investment by society in these science missions. 
Independent of the capabilities of future dedicated spacecraft tracking 
networks, there will always be situations where additional 
sensitivity can have a dramatic effect on the scientific value of a 
mission during short-term, high priority activities.  For this reason 
the ability to track scientific spacecraft with the SKA is important 
for maximizing the benefit that society gets from its over-all 
investment in scientific activities.

The same argument can be (and has been) used to justify observing 
time for radio astronomy experiments on dedicated spacecraft tracking 
antennas when their sensitivity, frequency coverage, or geographic 
locations provided a significant increase in the value of the 
observations.  Considering the cost of large radio arrays, whether 
intended primarily for radio astronomy or for spacecraft tracking, it 
seems only prudent to allow flexibility for the occasional use of all 
existing arrays combined.  The potential beneficiaries of this 
flexibility will include deep space missions of NASA, ESA, JAXA, FSA, 
and other space agencies.
Even though the SKA may be used only occasionally for spacecraft 
tracking support, these occasions are likely to be highly visible 
and to have great public appeal.

\section{TYPES OF SPACE MISSIONS} 

Planetary exploration missions operate over a huge range of distances 
and data rates.  Many are not constrained by telemetry downlink data 
rates, but often those at the greatest distances or carrying the most 
advanced sensors are.  Limited sensitivity for telemetry downlink can 
also have important indirect effects on mission design.  As an 
example, consider a mission to sample the atmosphere of one of the 
outer planets.  

The traditional way to design such a mission is to 
use an orbiting or fly-by spacecraft to relay data from the 
atmospheric probes (parachute or balloon suspended, gliders, etc.) to 
Earth.  But going into orbit usually requires a very large fuel 
payload, and flybys tend to stay within sight of any given location 
on a planet for only brief periods.  If scientific data from the 
atmospheric probes could be received directly on Earth, no data relay 
would be needed, a single-point failure for the whole mission would 
be eliminated, and data could be obtained from an arbitrary number of 
probes on the Earth-facing hemisphere for up to half the planetary 
rotation period.  Figure 1 illustrates the increase in data rates 
that could be obtained on Earth by using the SKA.  

\begin{figure}[h]
\centerline{
\includegraphics[angle=-90,width=3.25in]{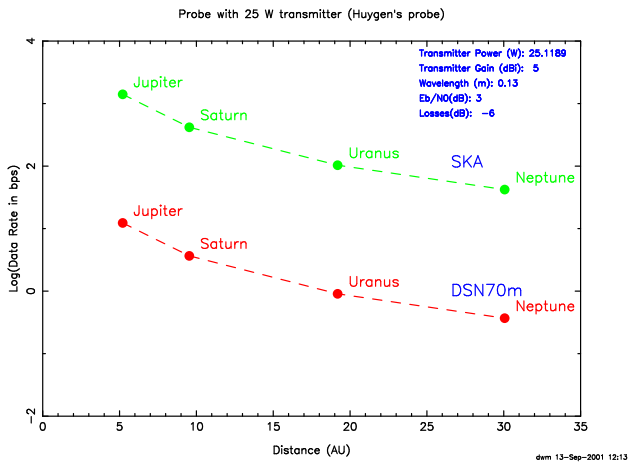} 
}
\caption{Telemetry downlink data rates to Earth from an 
atmospheric/surface probe similar to the Huygens probe on the 
Cassini mission, for each of the outer planets.  With the 
SKA, data rates up to a kb/s can be obtained, while the 
current DSN 70-m antennas can provide only a few 
bits/s at best.  Figure produced by D. Murphy at JPL.}   
\end{figure}

The potential benefits of SKA spacecraft tracking extend beyond 
simply obtaining more data from science missions of the type being 
flown today.  New mission concepts, such as large numbers of small, 
short-lived planetary probes (landers, rovers, or sub-surface penetrators), 
live video from cameras on planetary balloons or aircraft, 
constellations of hundreds of tiny in-situ magnetospheric probes, or 
solar-powered missions beyond the orbit of Mars can only be proposed 
if the on-board telecommunications hardware is extremely small, 
light, and low power.  More collecting area on the ground is the only 
way to make innovative missions like these possible.

An additional benefit of using the SKA is that the wide range of 
baseline lengths required for imaging also provides real-time 
astrometric position measurements using phase delays.  Phase delay 
is a more sensitive astrometric observable than the group delay 
currently used for spacecraft position measurement on single DSN 
baselines.  By combining phase-referencing techniques using 
in-beam reference sources (made possible by the high sensitivity 
of the SKA) and a large SNR spacecraft signal, it will be 
possible to routinely determine plane-of-sky spacecraft positions to 
better than 0.1 milliarcsecond.  This angular precision corresponds 
to a linear distance of less than one km at the distance of Saturn.

\section{CURRENT AND PREDICTED DSN CAPABILITIES} 

The most sensitive current facility for spacecraft tracking is the 
NASA Deep Space Network, which consists of one 70-m antenna and 
several 34-m antennas at each of three sites (California, Australia, 
and Spain).  These antennas are being upgraded to operate at 
frequencies of 32 GHz in addition to the traditional 2.3 and 8.4 GHz 
bands.
It has been recognized for some time that the sensitivity of the DSN 
needs to be increased by a large factor to properly support future 
missions.  In fact, even some current planetary missions can send back 
only a small fraction of the data obtained by their instruments due 
to the limited downlink data rates available.

Two options for increasing DSN downlink capability are being 
developed:  optical communications and large arrays of radio 
antennas.  Optical telecom has the potential to support extremely 
high data rates, but also has some serious disadvantages:  It 
requires highly accurate and stable pointing of the spacecraft 
optical system, which precludes use during critical entry, descent, 
and landing mission phases, or from small unstabilized platforms like 
balloons or parachute-supported atmosphere probes, or from any type 
of low gain antenna during a spacecraft emergency.  It also requires 
multiple large ground telescopes (for weather diversity) or dedicated 
orbiting telescopes for telemetry reception.

As a result of these difficulties, radio is likely to remain the 
dominant medium for spacecraft telemetry support for the foreseeable 
future.  In support of future radio telecom the DSN plans to build an 
array of approximately four hundred 12-m diameter antennas at each of 
their three sites during the next decade.  This will provide about an 
order of magnitude increase in sensitivity over the existing 70-m DSN 
antennas.  However, it is
worth noting that this will still be an order of magnitude below the 
sensitivity of the SKA.  Consequently even if the DSN arrays are 
built before the SKA, there will still be opportunities for the SKA 
to make very significant contributions to the scientific data return 
from high priority missions.

\section{REQUIREMENTS FOR TELEMETRY RECEPTION} 

The relevant SKA specifications to enable this type of observing are 
coverage of the major deep space telemetry downlink frequencies (2, 
8, and 32 GHz), coverage of the ecliptic region of the sky, and the 
availability of a beamformed (summed, with no time averaging) output 
signal from the inner part of the array.  Note that we never want to 
use the correlator output directly for telemetry reception.  We want 
a vector sum of the (properly aligned) antenna voltages as the basic 
output, not cross-products.  A continuous data stream is needed, 
since gaps in telemetry data cannot be recovered through additional 
observing.  This is a contrast with normal radio astronomy observing, 
where all data bits are equal!

There are two scenarios where use of the SKA may be justified - to 
support very high data rates that enhance the science return during 
short-duration mission phases (e.g., live video from an airplane in 
the Martian atmosphere), or to support low data rates from probes 
that could not otherwise send data directly to Earth at all (e.g., 
figure 1).  These two extremes require different operating modes. 
For high data rates the SNR of the spacecraft will be large, and 
real-time signal combining can be done with beamforming hardware 
alone using algorithms such as SUMPLE (see figure 2); there is no 
need for a full cross-correlator in this case.

\begin{figure}[h]
\includegraphics*[width=3.8in]{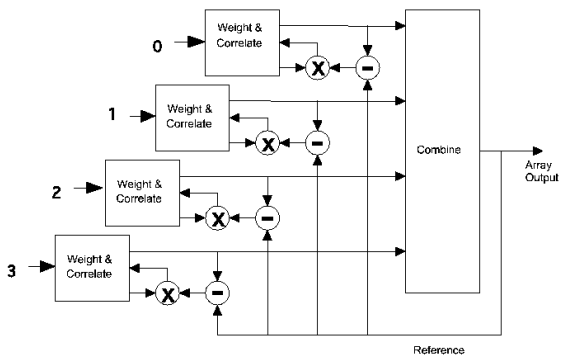}
\caption{SUMPLE architecture for spacecraft 
telemetry reception \cite{6}.  This is an iterative approach in which 
each antenna is correlated with the sum of the N-1 other antennas.  
Experience shows that stable delay and phase solutions are obtained 
rapidly.  No traditional N(N-1)/2 baseline cross-correlator 
is required. 
The architecture complexity scales linearly with the 
number of antennas being combined.  
} 
\end{figure}

For the low SNR case, it will be necessary to phase 
the inner part of the SKA using nearby radio 
sources and a correlator in the usual way, and supply the resulting 
antenna gain solutions to the beamformer (see figure 3).

\begin{figure}[h]
\includegraphics*[width=3.8in]{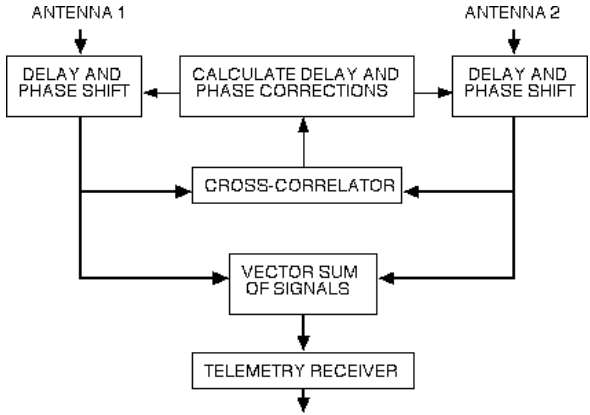}
\caption{One possible beamformer architecture for low-SNR  
spacecraft telemetry reception, adapted from \cite{5}.  An arbitrary 
number of antenna signals can be combined with this approach. 
Note that the correlator output is used only to determine delay 
and phase corrections needed for coherent signal summing.  The 
output is a vector sum of the input signal voltages, not a 
cross-correlation product. 
} 
\end{figure}

\section{REQUIREMENTS FOR ANGULAR TRACKING} 

The ability of connected-element interferometers to provide nearly 
instantaneous sky position measurements represents a new capability 
for spacecraft navigation.  Currently the plane-of-sky position of a 
spacecraft must be deduced from radial (Doppler and range) 
measurement using a model of the spacecraft trajectory (gravitational 
field of the solar system).  This requires relatively long tracking 
passes and is susceptible to errors in the model used. 
Alternatively, VLBI can be used to determine sky positions when a 
spacecraft transmits a special set of widely-spaced signals to allow 
group delay measurements. This special mode interrupts normal 
telemetry transmissions.

Unlike current spacecraft tracking networks that have only single 
very long baselines, the SKA will have a very large number of 
baselines covering a wide range of spacings.  Although motivated by 
imaging requirements, this type of array configuration also allows 
astrometric measurements based on phase delay instead of group delay. 
Phase delay measurements allow either greater astrometric precision, 
or similar precision with shorter baselines compared to group delay 
measurements.  The astrometric error from phase delay measurements is 
related to the ratio of baseline length to the observing frequency, 
while for group delay this is the baseline length over the bandwidth 
of the observations.  The SKA will have large fractional bandwidths, 
but this will not reduce the group delay error because the narrow 
spacecraft signal bandwidth will be the limiting factor.

Another advantage of a connected-element interferometer like the SKA 
is the speed with which positions measurements can be made.   This 
can be critical during the hours just before a spacecraft arrives at 
a target, particularly if aerocapture or atmospheric entry is going 
to be attempted.  In addition, the SKA will be able to subdivide or 
re-use its large collecting area, through sub-arraying or 
multibeaming (for phased array concepts).  This will permit all 
astrometric measurements to be made in a fully differential mode in 
which angularly nearby reference sources are used to correct errors 
caused by imperfect modeling of atmospheric and ionospheric delays or 
residual baseline offsets.  The one remaining significant error 
source for differential astrometry, phase offsets cause by partially 
resolved structure in the reference source(s), can be removed because 
the SKA will be able to produce good, high resolution images of any 
reference source it observes.

\section{UNIQUE ASPECTS OF SKA SUPPORT} 

In is likely that SKA support of science spacecraft will be 
infrequent but extremely valuable.  The main goal is to allow 
occasional significant increases in the quantity of (very expensively 
obtained) data from scientific space missions to be obtained.  The 
fact that the SKA can do this and simultaneously provide accurate 
plane-of-sky spacecraft positions for navigation is a unique new 
combination of capabilities.

The large A/T of the SKA can also be used for indirect support of 
science for space missions.  A clear example is atmospheric or ring 
occultation experiments, where the increased SNR provided by the SKA 
will allow useful data to be obtained over a much wider range of 
optical depths.

\begin{figure}[h]
\includegraphics*[width=3.7in]{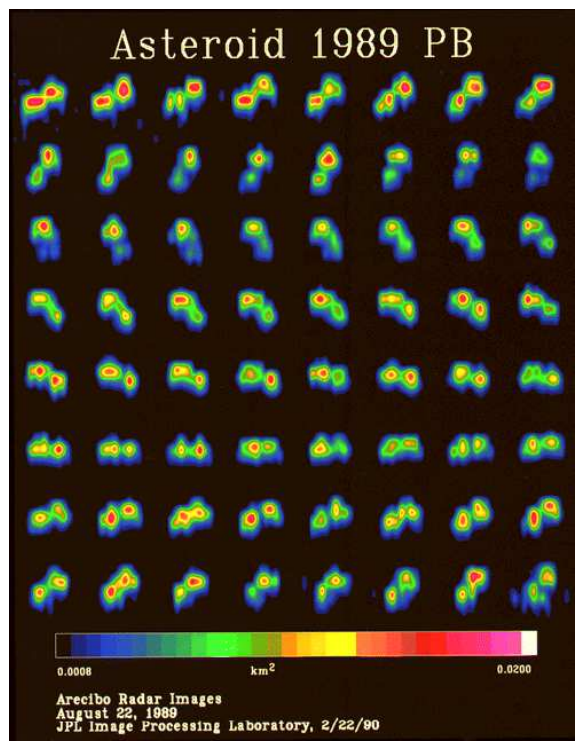} 
\caption{Radar images of an asteroid \cite{4}.  The resolution of 
images like these could be improved by
an order of magnitude or more by using the SKA to receive radar 
echos.} 
\end{figure}

Another area where ground-based collecting area could be particularly
valuable is planetary radar (see contribution by Butler, this volume).
While not a space mission, this scientific activity addresses many of
the same issues of planetary geology and evolution as flight projects
do.  Planetary radar experiments tend to be SNR-limited (except for
radar imaging of very-near-Earth asteroids), so an increase in
sensitivity for echo reception would be very valuable.  Targets
include planets and moons out to the distance of Saturn and Titan,
comets, and asteroids.  The frequencies used for planetary radar
observations are 2.3 and 8.5 GHz.

Radar measurements of the trajectories, sizes, and spin vectors 
of near Earth asteroids are essential for determining their future 
orbits accurately, and thus evaluating their potential harzard to 
Earth.  Figure 4 shows an example of delay-Doppler asteroid imaging 
with the Arecibo planetary radar system.  A bi-static radar system 
using the SKA for reception would significantly increase the 
available SNR of radar measurements.

The imaging capabilities of the SKA can be applied to radar 
observations by using the angular resolution of the SKA to resolve 
the ambiguities that plague traditional delay-Dopplar radar imaging. 
This will permit radar imaging with better angular resolution than 
the SKA can provide directly \cite{3}.

\section{EXAMPLES}

Examples of the enhanced science return that higher telemetry data 
rates can enable are easy to think of:  movies instead of single 
images, high resolution mapping and spectra instead of low 
resolution, mapping of entire planets instead of targeted 
observations.  At the other extreme, critical short-lived mission 
phases like atmospheric entry, descent, and landing can only use 
low-gain antennas because of rapid and unpredictable spacecraft 
motions.  In this situation the ability to receive even very small 
quantities of real-time data could be critical for diagnosing 
problems.  We have learned from past experience how damaging it can 
be to have no telemetry at all during a critical mission event that 
resulted in failure.  The same applies for a spacecraft emergency 
during any phase of a mission - the ability to get some data over 
low-gain spacecraft antennas can be immensely valuable for diagnosing 
problems and working out recovery procedures.

The flexibility of SKA observing implies that it should be possible 
to support spacecraft tracking and navigation in a co-observing mode. 
For example, while looking in the direction of a spacecraft, and 
array and correlator can be simultaneously used for large scale 
survey programs.  The correlator would operate in its normal imaging 
mode; all special-purpose signal processing would be done 
independently by a beamformer.

\section{CONCLUSIONS} 

The public invests in astronomy, and other scientific programs, for 
the excitement of exploration and discovery and for the chance of 
answering fundamental questions about the universe and our place in 
it.  Like astronomy, planetary exploration addresses very fundamental 
questions of broad public interest.  To maximize the scientific 
return from the significant investments in planetary science 
missions, we need to send as much data back to Earth as possible. 
This is particularly true for missions with short duration, high 
priority data-taking phases (e.g., atmospheric or surface probes on 
Venus, Europa, Io, or other hostile environments).  The SKA has the 
potential to greatly increase the data return from such missions. 
This is true independent of plans to build dedicated radio arrays for 
spacecraft tracking or for optical communications, as dedicated 
tracking arrays are unlikely to come close to the sensitivity of the 
SKA and optical communications are not possible with low gain 
antennas or through opaque atmospheres.  Occasional use of the SKA 
for telemetry reception during unique opportunities will contribute 
dramatically to the total increase in our scientific knowledge and to 
the public's interest in the SKA.

\end{document}